\documentclass[showpacs]{revtex4}
\usepackage{amsmath}
\usepackage{graphicx}
\usepackage{amssymb}
\usepackage{graphics}

\begin{document}

\title{Angular dependence of magnetoresistance and Fermi-surface shape in quasi-2D metals}
\author{P.D. Grigoriev}
\affiliation{L.D. Landau Institute for Theoretical Physics, Chernogolovka, Russia}
\date{\today}

\begin{abstract}
The analytical and numerical study of the angular dependence of
magnetoresistance in layered quasi-two-dimensional (Q2D) metals is
performed. The harmonic expansion analytical formulas for the angular
dependence of Fermi-surface cross-section area in external magnetic field are obtained for various typical crystal symmetries. 
The simple azimuth-angle dependence of the Yamaji angles is
derived for the elliptic in-plane Fermi surface. These formulas correct some
previous results and allow the simple and effective interpretation of the
magnetic quantum oscillations data in cuprate high-temperature
superconducting materials, in organic metals and other Q2D metals. 
The relation between the angular dependence of magnetoresistance 
and of Fermi-surface cross-section area is derived. The applicability region of all results obtained and of some previous widely used analytical results is investigated using the numerical calculations.
\end{abstract}
\maketitle

\section{Introduction}

The layered quasi-two-dimensional (Q2D) compounds attract great attention
for their novel physical properties and promising technical applications.
High-temperature cuprate superconductors,\cite{HTc} organic metals,\cite{IYS}
heterostructures,\cite{Heterostructures} intercalated graphites\cite%
{IntercalatedGraphitesReview2002} are the examples of these
compounds. The knowledge of quasiparticle dispersion in these
compounds is very important for understanding their properties and
electronic phase diagram. The traditional and powerful tools to
determine the Fermi surface (FS) geometry and the electron
dispersion in various metals are the magnetic quantum oscillations
(MQO)\cite{Shoenberg} and the angular dependence of
magnetoresistance (ADMR)\cite{MarkReview}. There is a huge amount
of publications, devoted to the experimental determination of the
FS geometry and electron dispersion in high-temperature cuprate
superconductors\cite{MQOHighTc,Bergemann}, in
MgB$_{2}$,\cite{MgB2MQOReview} in organic metals (see refs.
\cite{MarkReview},\cite{MQORev} for reviews) and in many other Q2D
metals. The interpretation of the MQO data is, usually, based on
the detailed comparison with the band-structure calculations,
which is a complicated and often ambiguous procedure. The
interpretation of ADMR is also based on fitting by the numerical
calculations with a large number of fitting
parameters.\cite{HusseyNature2003,AbdelPRL2007AMRO,McKenzie2007}
The quick and effective extraction of the FS geometry and of
electron dispersion from the experimental data on MQO and on ADMR
requires reliable and simple theoretical formulas.

The general form of the electron dispersion in Q2D \ compounds with
monoclinic or higher crystal symmetry can be expressed as the Fourier series
in cylindrical coordinates:%
\begin{equation}
\varepsilon \left( \mathbf{k}\right) =\sum_{\nu \geq 0,\mu =\text{even}%
}\epsilon _{\mu \nu }\left( k\right) \cos \left( \nu k_{z}c^{\ast }\right)
\cos \left( \mu \phi +\phi _{\mu \nu }\right) ,  \label{DispGen}
\end{equation}%
where the integers $\nu ,\mu \geq 0,$ the electron momentum $\mathbf{k}%
=\left( k_{x,}k_{y},k_{z}\right) $ (we put $\hbar =1$), $c^{\ast }$ is the
interlayer lattice constant, $k=\sqrt{k_{x}^{2}+k_{y}^{2}}$ is the absolute
value of the in-plane momentum and $\phi =\arctan \left( k_{y}/k_{x}\right) $
is the in-plane angle (i.e., the azimuth angle in spherical coordinates). In
triclinic crystals the only symmetry constraint on the electron dispersion
is $\varepsilon \left( \mathbf{k}\right) =\varepsilon \left( -\mathbf{k}%
\right) $, and the electron dispersion (\ref{DispGen}) may also contain the
additional terms
\begin{equation}
\Delta \varepsilon \left( \mathbf{k}\right) =\sum_{\nu >0}\sum_{\mu =\text{%
odd}}\epsilon _{\mu \nu }\left( k\right) \sin \left( \nu k_{z}c^{\ast
}\right) \cos \left( \mu \phi +\phi _{\mu \nu }\right) .  \label{OddDisp}
\end{equation}%
For simplicity, below we only consider the case of monoclinic or higher
crystal symmetry, where the terms (\ref{OddDisp}) are absent.

Usually, it is sufficient to keep only the first few terms in the infinite
series (\ref{DispGen}). For example, if the interlayer transfer integral of
conducting electrons, $t_{c}\left( \phi \right) \approx \epsilon _{01}$, is
much smaller than the in-plane band width $\epsilon _{00}$, the
tight-binding approximation can be used, and one keeps only the terms with $%
\nu =0$ and $\nu =1$:%
\begin{equation}
\varepsilon \left( \mathbf{k}\right) =\varepsilon \left( k,\phi \right)
-2t_{c}\left( \phi \right) \cos (k_{z}c^{\ast }).  \label{1}
\end{equation}%
The FS, being given by the equation $\varepsilon \left( \mathbf{k}\right)
=E_{F}$, is a warped cylinder in Q2D compounds. If magnetic field is applied
along the $z$-axis of the Q2D metals with the electron dispersion (\ref{1}),
there are two extremal FS cross-section areas $A_{ext}$ encircled by the
closed curves $\varepsilon \left( k_{x},k_{y}\right) \pm 2t_{c}$ $=E_{F}$.
Hence, the two close fundamental frequencies $F_{1,2}=\left( c/2\pi e\hbar
\right) A_{ext}$ appear in MQO, giving the beats of MQO.\cite{Shoenberg} The
temperature dependence of the MQO amplitude gives the cyclotron mass for the
extremal orbit: $m_{ext}^{\ast }\equiv \left( 1/2\pi \right) \left[ \partial
A_{ext}/\partial E\right] $. The beat frequency $\Delta F=F_{1}-F_{2}$ gives
the interlayer transfer integral: $4t_{c}=\Delta F\left( e\hbar
/m_{ext}^{\ast }c\right) $. The difference of the two extremal cross-section
areas $\Delta A_{ext}=2\pi e\hbar \,\Delta F/c$ and, hence, the beat
frequency depend on the magnetic field direction. In the first order in the
interlayer transfer integral and for the axially symmetric FS, this
dependence is given by\cite{Yam}
\begin{equation}
\Delta A_{ext}\propto J_{0}\left( c^{\ast }k_{F}\tan \theta \right) ,
\label{Yam}
\end{equation}%
where $J_{0}$ is the Bessel function, $k_{F}$ is the in-plane Fermi momentum
and $\theta $ is the tilt angle of magnetic field $\mathbf{B}$ with respect
to the $z$-axis (the polar angle of $\mathbf{B}$). Eq. (\ref{Yam}) was first
derived geometrically by Yamaji \cite{Yam} to explain the oscillating
angular behavior\cite{ExpAMROMark} of interlayer magnetoresistance in Q2D
organic metals. As the difference between the two extremal cross-section
areas is proportional to the interlayer transfer integral $t_{c}\left(
\theta \right) $, Eq. (\ref{Yam}) suggests that the interlayer transfer
integral has the similar angular dependence:%
\begin{equation}
t_{c}\left( \theta \right) \approx t_{c}\left( 0\right) J_{0}\left( c^{\ast
}k_{F}\tan \theta \right) ,  \label{tc}
\end{equation}%
which gives a strong angular dependence of interlayer magnetoresistance $%
\rho _{zz}\propto t_{c}^{2}\left( \theta \right) $. Eq. (\ref{tc}) was later
confirmed by the quantum-mechanical calculation of the amplitude of
interlayer electron tunnelling in tilted magnetic field.\cite{Kur} The
angles $\theta _{m}$, for which the Bessel function has zeros:
\begin{equation}
J_{0}\left( c^{\ast }k_{F}\tan \theta _{m}\right) =0,  \label{YamZero}
\end{equation}%
are called the Yamaji angles and used to determine the in-plane Fermi
momentum $k_{F}$. At these angles both the interlayer magnetoresistance and
the amplitude of MQO have maxima. Usually, the in-plane electron dispersion $%
\varepsilon \left( k_{x},k_{y}\right) $ is anisotropic, and Eqs. (\ref{Yam}%
),(\ref{tc}) acquire a $\varphi $-dependent correction, where $\varphi $ is
the azimuthal angle of the magnetic field direction, $\tan \varphi
=B_{y}/B_{x}$. There is a considerable practical need of the simple
analytical formula for the $\varphi $-dependence of AMRO and MQO, which can
be used to extract the in-plane electron dispersion from the experimental
data.

The widely used analytical result for the $\varphi $-dependence of the FS
cross section, derived by Bergemann et al.\cite{Bergemann} and given by Eq. (%
\ref{A1}) below, takes the FS corrugation only in the first order, which is
not enough to obtain correctly even the main $\varphi $-dependent term in
the angular dependence of the cross-section area. Another simple and widely
used\cite{Mark92,Nam1,HousePRB1996} analytical result for the $\varphi $%
-dependence of AMRO maxima (Yamaji angles),%
\begin{equation}
\tan \theta _{n}\approx \pi \left( n-1/4\right) /p_{B}^{\max }c^{\ast },
\label{DThetaS}
\end{equation}%
with $p_{B}^{\max }$ being the maximum value of the Fermi momentum
projection on the in-plane magnetic field direction, was derived\cite{Mark92}
from the Shockley tube integral\cite{Ziman} using the saddle point
approximation. This approximation assumes that the $z$-component of the
electron velocity oscillates rapidly when the electron moves along its
closed classical orbit in the momentum space in magnetic field. This is
valid only at high tilt angles $\theta $ of magnetic field, when $\tan
\theta \gg 1/c^{\ast }p_{B}^{\max }\left( \varphi \right) $, and only in the
very clean samples with $\omega _{c}\tau \cos \theta \gg 1$, where $\omega
_{c}$ is the cyclotron frequency and $\tau $ is the electron mean free time.
For the first Yamaji angle this derivation is too approximate because $%
c^{\ast }p_{B}^{\max }\left( \varphi \right) \tan \theta \sim 1$. Below we
show that Eq. (\ref{DThetaS}) is valid only for the elliptical FS in the
limit $\omega _{c}\tau \cos \theta \gg 1$. With some small error it can also
be applied to the FS, which is close to elliptical. However, Eq. (\ref%
{DThetaS}) gives completely wrong result for the $\varphi $-dependence of
Yamaji angles when the in-plane FS has tetragonal (as in cuprate high-Tc
superconductors) or hexagonal (as in MgB$_{2}$\cite{MgB2MQOReview} or
intercalated graphites\cite{IntercalatedGraphitesReview2002}) symmetry.

The aim of the present paper is to derive the new suitable analytical
formulas for the $\varphi $-dependence of the FS cross-section area, Yamaji
angles and magnetoresistance, which can be used to extract the FS parameters
from the experimental data. The applicability region of some previous and
widely used results will also be studied.

In Sec. II we write down the relation between the dispersion (\ref{DispGen})
and FS harmonic expansion. In Sec. III we find the main $\varphi $-dependent
correction to the FS cross-section area for the anisotropic dispersion $%
\varepsilon \left( k_{x},k_{y}\right) $\ in Eq. (\ref{1}), when the in-plane
anisotropy of the FS is weak. As will be shown, this result has wide
applicability region and can also be applied to almost square-shaped
in-plane FS as in the high-Tc cuprate superconductors. In Sec. IV we derive
the exact expression for the Yamaji zeros for the elliptical FS shape. The
deviations from this result for non-elliptic FS will also be studied. In
Sec. V we derive the relation between the $k_{z}$-dependence of the
cross-section area and magnetoresistance in the clean samples, where $\omega
_{c}\tau \cos \theta \gg 1$. This relation shows, that the geometrical and
resistivity Yamaji angles coincide in the limit $\omega _{c}\tau \cos \theta
\gg 1$. The discussion and summary of the results is given in Sec. VI.

\section{Fermi surface parametrization}

The dependence of the Fermi momentum $k_{F}\left( \phi ,k_{z}\right) $ on
the polar angle $\phi $ and the momentum component $k_{z}$ can be expanded
in the Fourier series:\cite{Bergemann}%
\begin{eqnarray}
k_{F}\left( \phi ,k_{z}\right) &=&\sum_{\nu \geq 0}k_{\nu }\left( \phi
\right) \cos \left( \nu k_{z}c^{\ast }\right)  \label{kF0} \\
&=&\sum_{\mu ,\nu \geq 0}k_{\mu \nu }\cos \left( \nu k_{z}c^{\ast }\right)
\cos \left( \mu \phi +\phi _{\mu }\right) .  \label{kF}
\end{eqnarray}%
The Fermi momentum satisfies$~$the equation
\begin{equation}
\varepsilon \left[ k_{F}\left( \phi ,k_{z}\right) ,\phi ,k_{z}\right] =E_{F},
\label{kFEq}
\end{equation}%
where $~E_{F}=\varepsilon \left( k_{F}\right) $ is the Fermi energy. The
coefficients $\epsilon _{\mu \nu }$ in Eq. (\ref{DispGen}) are related to
the coefficients $k_{\mu \nu }$ in the FS parametrization (\ref{kF}) through
the equation%
\begin{equation}
\sum_{\nu \geq 0,\mu =\text{even}}\epsilon _{\mu \nu }\left[ k_{F}\left(
\phi ,k_{z}\right) \right] \cos \left( \nu k_{z}c^{\ast }\right) \cos \left(
\mu \phi +\phi _{\mu \nu }\right) =E_{F}.  \label{EqFourrier}
\end{equation}%
This equation on $k_{\mu \nu }$ can be solved by the iteration procedure,
assuming that the warping coefficients $\epsilon _{\mu \nu }/\epsilon _{00}$
are small and fall down rapidly with increasing $\mu $ and $\nu $. In the
first order, each term $k_{\mu \nu }$ in the series (\ref{kF}) comes only
from the term $\epsilon _{\mu \nu }$ in (\ref{DispGen}) with the same
indices $\mu ,\nu $:%
\begin{equation}
k_{\mu \nu }^{\left( 1\right) }=-\epsilon _{\mu \nu }\left( k_{F}\right)
/\epsilon _{00}^{\prime }\left( k_{F}\right) .  \label{kmn}
\end{equation}%
In the second order in $\epsilon _{\mu \nu }$, the coefficients $k_{\mu \nu
} $ come from the interference of the infinite number of the terms $\epsilon
_{\mu ^{\prime }\nu ^{\prime }}$ and $\epsilon _{\mu ^{\prime \prime }\nu
^{\prime \prime }}$ in the dispersion (\ref{DispGen}), such that $\mu =\mu
^{\prime }\pm \mu ^{\prime \prime }$ and $\nu =\nu ^{\prime }\pm \nu
^{\prime \prime }$.

For simplicity, we take the dispersion (\ref{1}) and assume that the $k_{z}$
dependence of the energy is weak, i.e. the interlayer transfer integral $%
t_{c}\ll E_{F}$. The solution of equation%
\begin{equation*}
\varepsilon \left( k,\phi \right) =E_{F}+2t_{c}\cos (k_{z}c^{\ast }),
\end{equation*}
in the first order in the interlayer transfer integral $t_{c}=t_{c}\left(
\phi \right) ,$ gives the FS shape in the cylindrical coordinates:
\begin{equation}
k_{F}\left( \phi ,k_{z}\right) =k_{0}\left( \phi \right) +k_{1}\left( \phi
\right) \cos (k_{z}c^{\ast }),  \label{kF1}
\end{equation}%
where $k_{0}\left( \phi \right) $ satisfies $\varepsilon \left[ k_{0}\left(
\phi \right) ,\phi \right] =E_{F}$\ and
\begin{equation}
k_{1}\left( \phi \right) =2t_{c}\left( \phi \right) /\left[ \partial
\varepsilon \left( k,\phi \right) /\partial k\right] |_{k=k_{0}\left( \phi
\right) }.  \label{k1}
\end{equation}%
The partial derivative $\left( \partial \varepsilon /\partial k\right)
|_{k=k_{0}\left( \phi \right) }$ is the projection of the Fermi velocity on
the line, connecting the point on the FS with the coordinate origin $\mathbf{%
k}=0$. It depends on the electron dispersion and on the azimuthal angle $%
\phi $.

For the quite general form of the electron dispersion,
\begin{equation}
\varepsilon \left( k,\phi \right) =k^{\alpha }g\left( \phi \right) ,
\label{e1}
\end{equation}%
where $g\left( \phi \right) $ is an arbitrary function and $\alpha $ is also
arbitrary, the derivative
\begin{equation}
\left[ \partial \varepsilon \left( k,\phi \right) /\partial k\right]
|_{k=k_{0}\left( \phi \right) }=E_{F}/k_{0}\left( \phi \right) .  \label{dep}
\end{equation}%
The superelliptic dispersion
\begin{equation}
\varepsilon \left( k_{x},k_{y}\right) =\left( k_{x}/k_{1}\right) ^{\alpha
}+\left( k_{y}/k_{2}\right) ^{\alpha },  \label{SupElDisp}
\end{equation}%
which includes both linear and quadratic dispersions, is only a particular
case of the dispersion (\ref{e1}). With the relation (\ref{dep}), Eq. (\ref%
{kF1}) simplifies to%
\begin{equation}
k_{F}\left( \phi ,k_{z}\right) =k_{0}\left( \phi \right) \left[ 1+\frac{%
2t_{c}\left( \phi \right) }{E_{F}}\cos (k_{z}c^{\ast })\right] .
\label{kF1s}
\end{equation}%
However, the relation (\ref{dep}) may violate in some compounds, and the
application of the simplified formula (\ref{kF1s}) instead of Eqs. (\ref{kF1}%
),(\ref{k1}) requires additional proof.

\section{Cross-section area}

If the magnetic field is applied at polar and azimuthal angles $\theta $ and
$\varphi $,\ the Fermi surface cross-sectional area $A=A\left( k_{z0},\theta
,\varphi \right) $, cutting the $z$-axis at $k_{z}=k_{z0}$ and perpendicular
to the field, is given by the integral
\begin{equation}
A\left( k_{z0},\theta ,\varphi \right) =\int_{0}^{2\pi }d\phi ^{\prime
}k_{F}^{2}\left( \varphi +\phi ^{\prime },k_{z}\right) /\left( 2\cos \theta
\right) ,  \label{A}
\end{equation}%
where $\phi ^{\prime }$ is the angle in the $x$-$y$ plane between the
direction of magnetic field and a point on the FS, and $k_{z}$ at this FS
point satisfies the equation
\begin{equation}
k_{z}=k_{z0}-k_{F}\left( \varphi +\phi ^{\prime },k_{z}\right) \tan \theta
\cos \phi ^{\prime }.  \label{kappa}
\end{equation}

Eqs. (\ref{A}) and (\ref{kappa}) allow to find the cross-section area $%
A\left( k_{z0},\theta ,\varphi ,k_{F}c^{\ast }\right) $ numerically for any
given FS, determined by the function $k_{F}\left( \phi ,k_{z}\right) $ or,
equivalently, by the coefficients $k_{\mu \nu }$ in the expansion (\ref{kF}%
). In practice, one usually solves the inverse problem of the extraction of
FS parameters from the experimental data on MQO or AMRO. Then, the direct
procedure of fitting the experimental data by the parameters in the
expansion (\ref{kF}) is rather ambiguous because of too large number of
fitting parameters. Usually, the coefficients $k_{\mu \nu }$ fall down
rapidly with increasing $\mu $ and $\nu $. Therefore, it is useful to fit
only the first few terms in the similar harmonic expansion of the
cross-section area
\begin{equation}
A\left( k_{z0},\theta ,\varphi \right) =\sum_{\mu ,\nu }A_{\mu \nu }\left(
\theta \right) \cos \left[ \mu \varphi +\delta _{\mu }\right] \cos \left(
\nu c^{\ast }k_{z0}\right) ,  \label{Aexp}
\end{equation}%
keeping only the first few terms in the expansion (\ref{kF}). The first
coefficients $A_{\mu \nu }\left( \theta \right) $ can be found analytically
in the main order in $k_{\mu \nu }$. The analytical formula for the
coefficients $A_{\mu \nu }\left( \theta \right) $ of the cross-section area
is especially useful because of their rather complicated dependence on $%
\theta $. In Sec. V it will be shown that the coefficient $A_{\mu 1}\left(
\theta \right) $ is directly related to the angular dependence of
magnetoresistance at $\omega _{c}\tau \gg 1$ and $t_{c}/E_{F}\ll 1$.

In the zeroth order in coefficients $k_{\mu \nu }$ in the expansion (\ref{kF}%
), i.e. for cylindrical FS neglecting any warping and in-plane asymmetry,
one obtains the trivial result $A^{\left( 0\right) }=\pi k_{F}^{2}/\cos
\theta $, where $k_{F}\equiv k_{00}$. In the first order in these
coefficients $k_{\mu \nu }$, one can neglect the dependence $k_{F}\left(
\phi ,k_{z}\right) $ in Eq. (\ref{kappa}) and substitute $%
k_{z}=k_{z0}-k_{F}\tan \theta \cos \phi ^{\prime }$ to Eq. (\ref{kF}). Then,
substituting Eq. (\ref{kF}) to Eq. (\ref{A}), one obtains the first order
correction to $A^{\left( 0\right) }$:
\begin{gather*}
A^{\left( 1\right) }\left( k_{z0},\theta ,\varphi \right) =\int_{0}^{2\pi
}d\phi ^{\prime }\frac{k_{F}^{2}\left( \varphi ,\phi ^{\prime
},k_{z0}\right) -k_{00}^{2}}{2\cos \theta } \\
\approx \int_{0}^{2\pi }\frac{k_{00}d\phi ^{\prime }}{\cos \theta }\sum_{\mu
,\nu \geq 0}{}^{\prime }k_{\mu \nu }\cos \left[ \mu \left( \varphi +\phi
^{\prime }\right) +\phi _{\mu }\right] \\
\times \cos \left[ \nu \left( k_{z0}-k_{F}\tan \theta \cos \phi ^{\prime
}\right) c^{\ast }\right] ,
\end{gather*}%
where the sum $\sum_{\mu ,\nu \geq 0}{}^{\prime }$ does not include the term
$\mu =\nu =0$. Since $\mu $ is even, one can replace in the integrand (here
and later we introduce the notation $\kappa \equiv k_{F}c^{\ast }\tan \theta
$)
\begin{equation}
\cos \left[ \nu \left( k_{z0}-k_{F}\tan \theta \cos \phi ^{\prime }\right)
c^{\ast }\right] \rightarrow \cos \left[ \nu k_{z0}c^{\ast }\right] \cos %
\left[ \nu \kappa \cos \phi ^{\prime }\right] .  \label{sub1}
\end{equation}%
One can also replace in the integrand
\begin{equation}
\cos \left[ \mu \left( \varphi +\phi ^{\prime }\right) +\phi _{\mu }\right]
\rightarrow \cos \left[ \mu \varphi +\phi _{\mu }\right] \cos \left[ \mu
\phi ^{\prime }\right] ,  \label{sub2}
\end{equation}%
because all odd terms vanish after the integration over $\phi ^{\prime }$.
Then, after the integration over $\phi ^{\prime }$, the correction $%
A^{\left( 1\right) }$ in the first-order in $k_{\mu \nu }$ writes down as
\begin{equation}
A^{(1)}=\frac{2\pi k_{00}}{\cos \theta }\sum_{\mu ,\nu \geq 0}{}^{\prime
}(-1)^{2\mu }\,k_{\mu \nu }\cos \left[ \mu \varphi +\phi _{\mu }\right] \cos
\left( \nu k_{z0}c^{\ast }\right) J_{\mu }\left( \nu \kappa \right)
\label{A1}
\end{equation}%
in agreement with Eq. (2) of Ref. \cite{Bergemann}. Since $J_{\mu }\left(
0\right) =0$ for $\mu \neq 0$, all terms $\sim k_{\mu 0}$ vanish in (\ref{A1}%
). This is natural, because in the zeroth order in $t_{c}$ the cross-section
area
\begin{equation}
A^{\left( 0\right) }\left( k_{z0},\theta ,\varphi \right) =\int_{0}^{2\pi
}d\phi ^{\prime }\frac{k_{0}^{2}\left( \varphi +\phi ^{\prime }\right) }{%
2\cos \theta }  \label{A0}
\end{equation}%
is independent of $\varphi $. Hence, to extract any information about the $%
\varphi $-dependence of the FS, one needs to consider the first order in $%
t_{c}/E_{F}$, i.e. to find $A_{\mu 1}\left( \theta \right) $. Thus, the $%
\varphi $-dependence of the cross-section area starts from the term $k_{\mu
1}$, which is of the same order as the second order term $k_{\mu
0}k_{01}/k_{F}$ [see Eq. (\ref{kF1s})]. Since Eq. (\ref{A1}), or Eq. (2) in
Ref. \cite{Bergemann}, is derived only in the first order in $k_{\mu \nu }$,
it does not give the correct $\varphi $-dependence of the cross-section area
even in the lowest $\varphi $-dependent order. This is illustrated below in
Figs. \ref{FigMon},\ref{FigTetr}. The extraction of the higher harmonics
using Eq. (\ref{A1}) is even more incorrect.

Let us calculate more accurately the lowest-order $\varphi $-dependent term
in the cross-section area, which is given by the coefficient $A_{\mu
1}\left( \theta \right) $ in the Fourier expansion. To calculate this
coefficient in the main order in FS warping, it is sufficient to use the FS
shape in the first order in $t_{c}$, given by Eq. (\ref{kF1}). Then, in the
same order, Eq. (\ref{A}) rewrites
\begin{equation}
A\left( k_{z0},\theta ,\varphi \right) \approx \int_{0}^{2\pi }d\phi
^{\prime }\frac{k_{0}^{2}\left( \varphi +\phi ^{\prime }\right) }{2\cos
\theta }\left[ 1+\frac{2k_{1}\left( \varphi +\phi ^{\prime }\right) }{%
k_{0}\left( \varphi +\phi ^{\prime }\right) }\cos (k_{z}c^{\ast })\right] ,
\label{A1i}
\end{equation}%
and substituting Eqs. (\ref{kappa}) and (\ref{sub1}), we obtain the
following expression for correction to $A\left( k_{z0},\theta ,\varphi
\right) $:
\begin{equation}
A^{\left( 1\right) }=\frac{\cos \left[ c^{\ast }k_{z0}\right] }{\cos \theta }%
\int_{0}^{2\pi }d\phi ^{\prime }k_{0}\left( \phi ^{\prime }\right)
k_{1}\left( \phi ^{\prime }\right) \cos \left[ c^{\ast }k_{0}\left( \phi
^{\prime }\right) \tan \theta \cos \left( \phi ^{\prime }-\varphi \right) %
\right] .  \label{A1N}
\end{equation}%
Here we have also changed the integration variable: $\phi ^{\prime
}\rightarrow \phi ^{\prime }+\varphi $. \ The Yamaji formula (\ref{Yam}) is
easily obtained from (\ref{A1N}) after taking $k_{0}\left( \phi \right)
=k_{F}=const,~k_{1}\left( \phi \right) =C_{1}\left( 2t_{c}/E_{F}\right)
k_{F}=const,$ where the dispersion-dependent constant
\begin{equation}
C_{1}\equiv \left( E_{F}/k_{F}\right) /\left( \partial \varepsilon
_{00}/\partial k\right) |_{k=k_{F}}\sim 1,  \label{C1}
\end{equation}%
and the integration over $\phi ^{\prime }$, resulting to%
\begin{equation}
A_{01}\left( \theta \right) =\frac{2t_{c}}{E_{F}}C_{1}\frac{2\pi k_{F}^{2}}{%
\cos \theta }J_{0}\left( \kappa \right) .  \label{A01}
\end{equation}%
The lowest-order $\varphi $-dependence of the cross-section area is
determined by the Fourier coefficient $A_{m1}\left( \theta \right) $, given
by%
\begin{equation}
A_{m1}\left( \theta \right) =\int_{0}^{2\pi }\frac{\cos \left( m\varphi
+\varphi _{m1}\right) d\varphi }{\pi \cos \theta }\frac{A^{\left( 1\right)
}\left( k_{z0},\theta ,\varphi \right) }{\cos \left[ c^{\ast }k_{z0}\right] }%
.
\end{equation}%
Performing the integration over $\varphi $, we obtain
\begin{equation}
A_{m1}\left( \theta \right) =\frac{2\left( -1\right) ^{m/2}}{\cos \theta }%
\int_{0}^{2\pi }d\phi ^{\prime }k_{0}\left( \phi ^{\prime }\right)
k_{1}\left( \phi ^{\prime }\right) \cos \left( m\phi ^{\prime }+\varphi
_{m1}\right) J_{m}\left[ c^{\ast }k_{0}\left( \phi ^{\prime }\right) \tan
\theta \right] .  \label{Am1}
\end{equation}%
To go further, we need to specify the functions $k_{0}\left( \phi \right) $
and $k_{1}\left( \phi \right) $. We distinguish two symmetries of electron
dispersion, namely, with straight and $\varphi $-dependent (corrugated in
the main order) interlayer transfer integral.

\subsection{Straight interlayer hopping}

When the in-plane FS anisotropy is weak, one can keep only the first $\phi $%
-dependent term in the Fourier expansion of the functions $k_{0}\left( \phi
\right) $ and $k_{1}\left( \phi \right) $. If the crystal symmetry allows
the $\phi $-independent (straight) interlayer coupling, these functions
expand as
\begin{eqnarray}
k_{0}\left( \phi \right) &\approx &\left( 1+\beta \cos m\phi \right) k_{F},
\label{kphim} \\
k_{1}\left( \phi \right) &\approx &\frac{2t_{c}C_{1}}{E_{F}}\left( 1+\beta
_{1}\cos m\phi \right) k_{F},  \notag
\end{eqnarray}%
where $m$ is an even integer number, $\left\vert \beta \right\vert
,\left\vert \beta _{1}\right\vert \ll 1$ and the constant $C_{1}\sim 1$ is
given by Eq. (\ref{C1}). Now we expand $J_{m}\left[ c^{\ast }k_{0}\left(
\phi ^{\prime }\right) \tan \theta \right] $ in the small parameter $\beta $
up to the first order (for the first Yamaji angle $\kappa \equiv c^{\ast
}k_{F}\tan \theta \approx 2.4\sim 1$), and the integral over $\phi ^{\prime
} $ in Eq. (\ref{Am1}) simplifies to%
\begin{eqnarray*}
&&\int_{0}^{2\pi }d\phi ^{\prime }\left( \beta +\beta _{1}\right) \cos
\left( m\phi ^{\prime }\right) \cos \left( m\phi ^{\prime }\right)
J_{m}\left( \kappa \right) \\
&&+\int_{0}^{2\pi }d\phi ^{\prime }\cos \left( m\phi ^{\prime }\right)
J_{m}^{\prime }\left( \kappa \right) \kappa \beta \cos \left( m\phi ^{\prime
}\right) \\
&=&\pi \beta \left\{ J_{m}\left( \kappa \right) \left( 1+\beta _{1}/\beta
\right) +J_{m}^{\prime }\left( \kappa \right) \kappa \right\} ,
\end{eqnarray*}%
where the derivative%
\begin{equation*}
J_{m}^{\prime }\left( \kappa \right) =\frac{dJ_{m}\left( \kappa \right) }{%
d\kappa }=\frac{m}{\kappa }J_{m}\left( \kappa \right) -J_{m+1}\left( \kappa
\right) .
\end{equation*}%
Hence, in the first order in $\beta $ we obtain
\begin{equation}
A_{m1}\left( \theta \right) =\left( -1\right) ^{m/2}\frac{4\pi
k_{F}^{2}C_{1}\beta t_{c}}{E_{F}\cos \theta }\left[ J_{m}\left( \kappa
\right) \left( 1+\frac{\beta _{1}}{\beta }+m\right) -\kappa J_{m+1}\left(
\kappa \right) \right] .  \label{Am1S}
\end{equation}%
Combining the results (\ref{Aexp}),(\ref{A01}) and (\ref{Am1S}), we obtain
the cross-section area
\begin{gather}
A\left( k_{z0},\theta ,\varphi \right) \approx \frac{\pi k_{F}^{2}}{\cos
\theta }+\frac{4\pi k_{F}^{2}t_{c}C_{1}}{E_{F}\cos \theta }\cos \left[
c^{\ast }k_{z0}\right] \times  \label{A1f} \\
\times \left\{ J_{0}\left( \kappa \right) +\beta \left( -1\right) ^{m/2}
\left[ \left( 1+\beta _{1}/\beta +m\right) J_{m}\left( \kappa \right)
-\kappa J_{m+1}\left( \kappa \right) \right] \cos \left( m\varphi \right)
\right\} .  \notag
\end{gather}%
The ratio $\beta _{1}/\beta $, entering this formula, can also be expressed
via the FS parametrization, given by Eq. (\ref{kF}): $\beta \equiv k_{\mu
0}/k_{00},~\beta _{1}=k_{\mu 1}/k_{01}$, $\beta _{1}/\beta =k_{\mu
1}k_{00}/k_{01}k_{\mu 0}$. The constant $C_{1}$ is equivalent to the
renormalization of $t_{c}$ and does not influence the Yamaji angles.
However, it changes the beat frequency of the magnetic quantum oscillations.
It also changes the amplitude of the $\varphi $-dependent term in the
cross-section area. For the dispersion of the form (\ref{e1}), i.e. if the
relation (\ref{dep}) satisfies, the ratio $\beta _{1}/\beta =1$ and the
constant $C_{1}=1$. For arbitrary dispersion, $\beta _{1}/\beta \sim 1$ and $%
C_{1}\sim 1$.
\begin{figure}[tbh]
\includegraphics[width=0.49\textwidth]{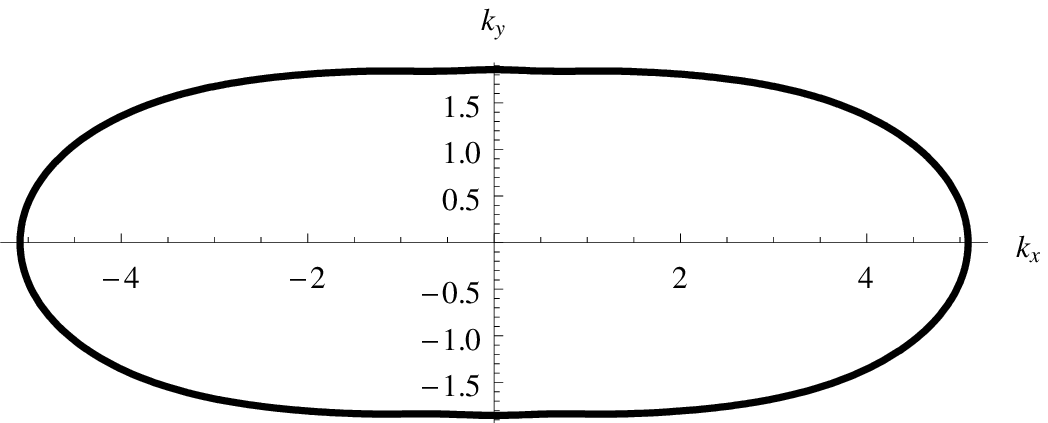}a\newline
\includegraphics[width=0.49\textwidth]{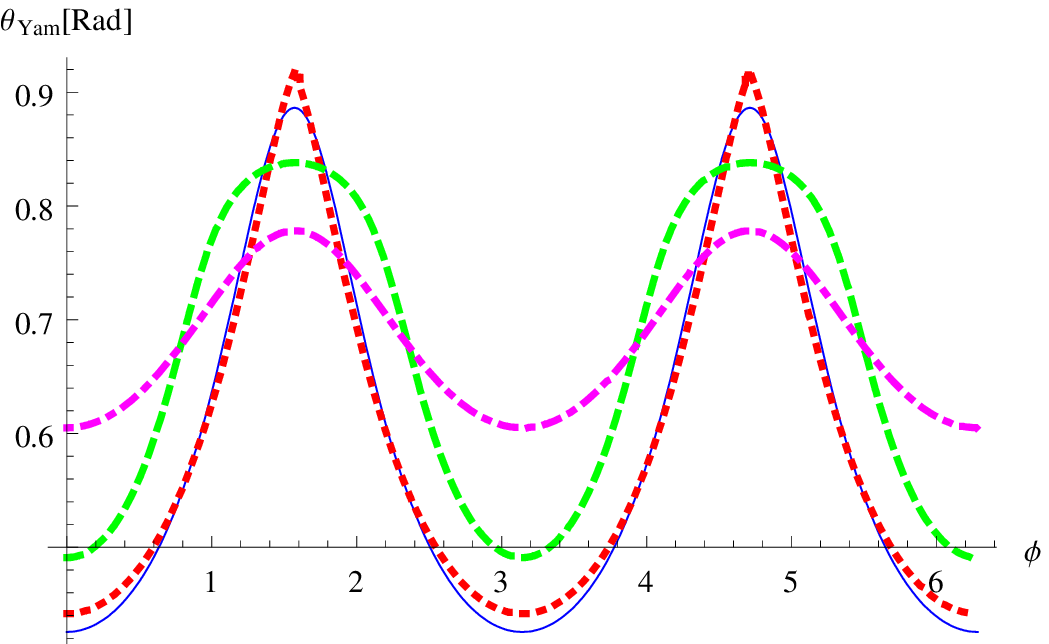}b
\caption{{}(Color online) The in-plane FS shape (a) and the first
Yamaji angle as function of the azimuth angle $\protect\varphi $
(b) for the
elongated FS with monoclinic symmetry and the Fermi momentum given by $%
k_{0}\left( \protect\phi \right) =\sum_{j=0}^{4}k_{4j\,0}\cos \left( 4j%
\protect\phi \right) $ with $\protect\beta =k_{40}/k_{00}=0.5$.
The higher harmonics are added to smooth the FS (see Fig. a). In
Fig. b the blue solid curve is the numerical result for the first
Yamaji angle, obtained using Eq.
(\protect\ref{A1N}); the red dotted curve is the result of Eq. (\protect\ref%
{YamZerosPhi}); the green dashed curve is the result from Eq. (\protect\ref%
{A1f}) and magenta dash-dotted curve is the result form Eq. (\protect\ref{A1}%
) derived by Bergemann et al.\protect\cite{Bergemann} One can see that Eq. (%
\protect\ref{YamZerosPhi}) gives the best result for Yamaji angles
when the FS is strongly elongated. The harmonic expansion still
gives a reasonable result, though the coefficient $\protect\beta
=0.5\sim 1$ is not small. The
result of Eq. (\protect\ref{A1}) gives much weaker $\protect\varphi $%
-dependence, than the correct one.} \label{FigMon}
\end{figure}

\begin{figure}[hhht]
\includegraphics[width=0.4\textwidth]{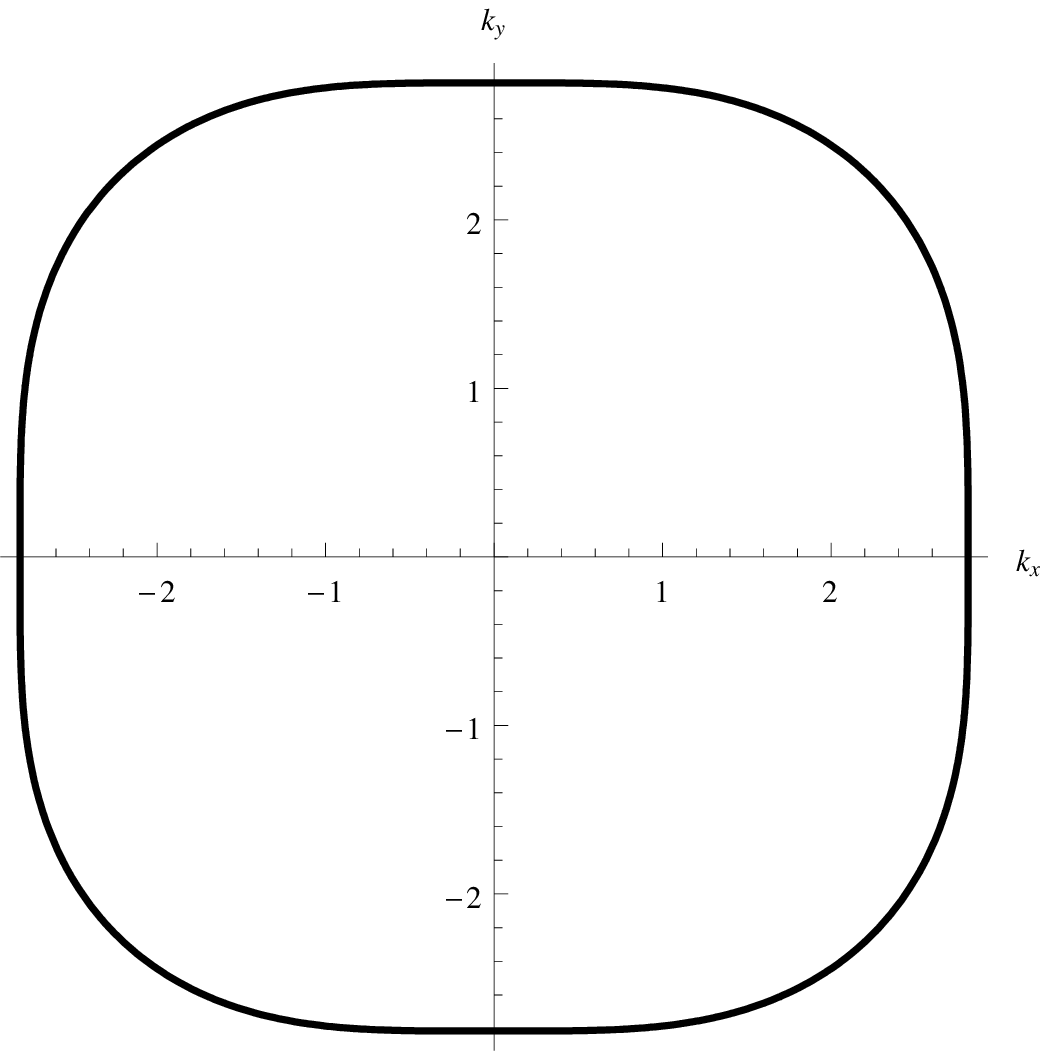}a\newline
\includegraphics[width=0.49\textwidth]{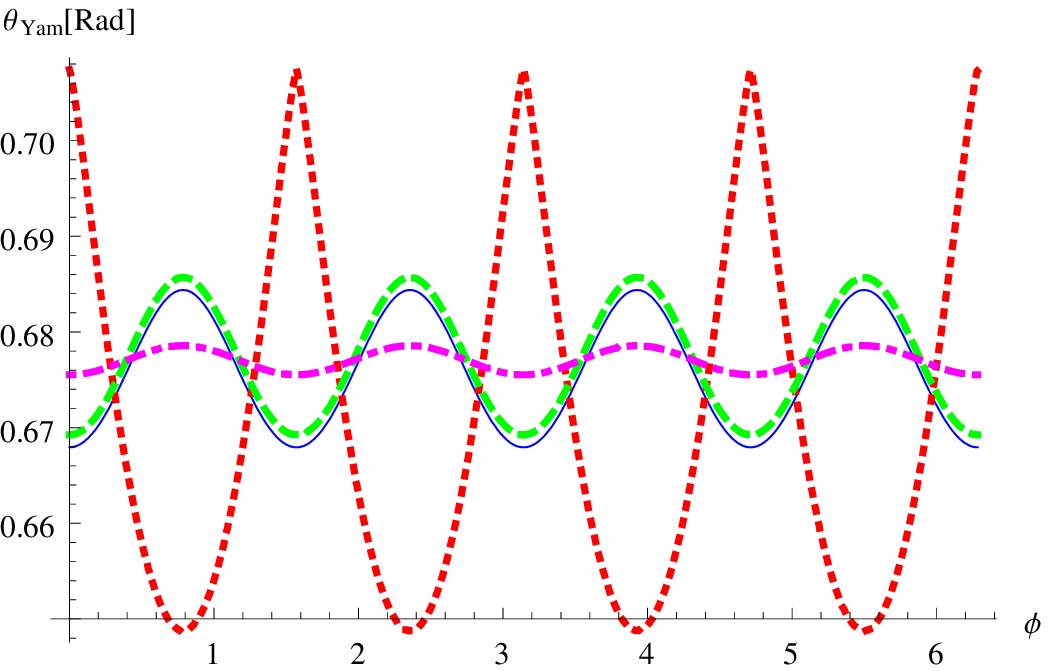}b
\caption{{}(Color online) The in-plane FS shape (a) and the first
Yamaji angle as function of the polar angle $\protect\phi $ (b)
for the FS with tetragonal symmetry and the Fermi momentum given
by Eq. (\ref{kphim}) with $\beta =-0.06$. Fig. a shows, that even
for very small parameter $\protect\beta $ the FS is strongly
different from the cylinder. It is almost quadratic. Hence,
keeping only the first term in the harmonic expansion of the
$\protect\phi $-dependence and the assumption that this term is
small, is applicable for most compounds with tetragonal symmetry.
In Fig. b the blue solid curve is the numerical result for the
first Yamaji angle, obtained using Eq. (\ref{A1N}). The red dashed
curve is the result of Eq. (\ref{YamZerosPhi}); it gives too
strong and opposite $\phi $-dependence, which is completely
incorrect. The green dotted curve is the result from Eq.
(\protect\ref{A1f}); it gives very good agreement with the
numerical result. The magenta dash-dotted curve illustrates Eq.
(\protect\ref{A1}), used by Bergemann et al.\cite{Bergemann}; this
result gives too weak $\phi $-dependence in agreement with the
discussion in the end of Sec. IIIA.} \label{FigTetr}
\end{figure}

The difference between two analytical results, given by Eqs. (\ref{A1f}) and
(\ref{A1}), is very strong: the factor $J_{m}\left( \kappa \right) $ in Eq. (%
\ref{A1}) is replaced by the completely different factor $\left[ \left(
1+\beta _{1}/\beta +m\right) J_{m}\left( \kappa \right) -\kappa
J_{m+1}\left( \kappa \right) \right] $ in Eq. (\ref{A1f}). First, the $%
\varphi $-dependence of the cross-section area, predicted by Eq. (\ref{A1f}%
), is stronger approximately by a factor $2+m$ than that of Ref. \cite%
{Bergemann}. Second, it may have different $\theta $-dependence due to the $%
J_{m+1}\left( \kappa \right) $ term, especially for high tilt angles $\kappa
\gtrsim 1$. To illustrate the above statement, we plot the results of Eqs. (%
\ref{A}), (\ref{A1f}), (\ref{A1}) and (\ref{YamZerosPhi}) in Figs. \ref%
{FigMon} and \ref{FigTetr} for the dispersions with monoclinic
$m=2$ and tetragonal $m=4$ symmetries.

\subsection{Strongly $\protect\phi $-dependent interlayer hopping}

In some compounds, e.g. in the high-temperature superconductors Sr$_{2}$RuO$%
_{4}$ and Tl$_{2}$Ba$_{2}$CuO$_{6+\delta }$,\cite{Bergemann,HusseyNature2003,AbdelPRL2007AMRO,McKenzie2007} the
body-centered tetragonal symmetry of the crystal leads to the $\phi $%
-dependent lowest-order interlayer transfer integral, $t_{c}\left( \phi
\right) =t_{c0}\sin \left( 2\phi \right) $, in the all or some parts of the
FS. Then, instead of Eq. (\ref{kphim}), we have
\begin{eqnarray}
k_{0}\left( \phi \right) &\approx &\left( 1+\beta \cos 2m\phi \right) k_{F},
\label{ko0} \\
k_{1}\left( \phi \right) &\approx &\frac{2t_{c}}{E_{F}}k_{F}C_{1}\sin \left(
m\phi \right) \left( 1+\beta _{1}\cos 2m\phi \right) .  \notag
\end{eqnarray}%
Substituting this into Eq. (\ref{Am1}) we obtain the main $\varphi $%
-dependent term, determined by the coefficient%
\begin{equation}
A_{m1}\left( \theta \right) \approx \frac{2\pi k_{F}^{2}\left( -1\right)
^{m/2}}{\cos \theta }\frac{2t_{c}C_{1}}{E_{F}}J_{m}\left( \kappa \right)
\label{Aom}
\end{equation}%
in agreement with the first-order result, given by Eq. (\ref{A1}). This term
does not depend on the in-plane FS anisotropy $\beta $. To extract this
anisotropy in the first order in the $\varphi $-dependent interlayer
transfer integral, one needs to consider $3m$ harmonic in the cross-section
area. For this we replace in Eq. (\ref{Am1}) $m\rightarrow 3m$, substitute
Eq. (\ref{ko0}) and perform the calculation, similar to that in the
derivation of Eq. (\ref{Am1S}). Then, in the lowest order in $\beta ,\beta
_{1}$ we obtain%
\begin{equation}
A_{3m~1}\left( \theta \right) =\frac{\left( -1\right) ^{3m/2}}{\cos \theta }%
\frac{2\pi t_{c}C_{1}}{E_{F}}\beta k_{F}^{2}\left[ \left( 1+\frac{\beta _{1}%
}{\beta }+3m\right) J_{3m}\left( \kappa \right) -\kappa J_{3m+1}\left(
\kappa \right) \right] .  \label{A3m}
\end{equation}%
This result differs from Eq. (\ref{Am1S}) by the replacement $m\rightarrow
3m $ (note, that $m=2$ for Sr$_{2}$RuO$_{4}$ and Tl$_{2}$Ba$_{2}$CuO$%
_{6+\delta }$), and the prefactor before the square brackets is two times
smaller. The difference between the first-order result of Eq. (\ref{A1}) and
the new formula (\ref{A3m}) for the $3m$ harmonic is even stronger than in
the case of Eq. (\ref{Am1S}). The total $\varphi $-dependence of the
cross-section area in the case of $\varphi $-dependent interlayer coupling,
given by Eq. (\ref{ko0}), writes down as
\begin{gather}
A\left( k_{z0},\theta ,\varphi \right) \approx \frac{\pi k_{F}^{2}}{\cos
\theta }+\frac{4\pi k_{F}^{2}t_{c}C_{1}}{E_{F}\cos \theta }\cos \left[
c^{\ast }k_{z0}\right] \times  \label{A1fo} \\
\times \left\{ J_{m}\left( \kappa \right) \sin \left( m\varphi \right) +%
\frac{\beta }{2}\left( -1\right) ^{3m/2}\left[ \left( 1+\frac{\beta _{1}}{%
\beta }+3m\right) J_{3m}\left( \kappa \right) -\kappa J_{3m+1}\left( \kappa
\right) \right] \sin \left( 3m\varphi \right) \right\} .  \notag
\end{gather}%
This formula can be applied to analyze the experimental data in
high-temperature superconductors Sr$_{2}$RuO$_{4}$, Tl$_{2}$Ba$_{2}$CuO$%
_{6+\delta }$, where $m=2$ in Eq. (\ref{A1fo}), and to some other layered
compounds with the appropriate symmetry. Note, that Eqs. (\ref{A1f}) and (%
\ref{A1fo}) were derived under condition $\beta \kappa \ll 1$, which is
fullfilled in the compounds of tetragonal or hexagonal symmetry at not very
high tilt angle of magnetic field. At very high tilt angle, $\tan \theta
>E_{F}/t_{c}$, above derivations are not valid also because of the multiple
intersections of the FS by the cross-section plane.

\subsection{Analysis of magnetic quantum oscillations}

The well-resolved magnetic quantum oscillations in quasi-2D metals give two
close frequencies $F_{\max }$ and $F_{\min }$, corresponding to the maximum
and minimum of the FS\ cross-section area. To extract the $\varphi $%
-dependence of the FS, as follows from Eqs. (\ref{A1f}) and (\ref{A1fo}),
one needs to measure the $\varphi $-dependence of the difference $\Delta
F=F_{\max }-F_{\min }\sim 4\pi k_{F}^{2}t_{c}/E_{F}$ between these two close
frequencies (the beat frequency), which is harder because requires the
resolution of MQO in the wider interval of magnetic field. The observation
of the beat frequency itself is important, because it means the existence of
the 3D Fermi surface, i.e. of the coherent interlayer electron transport.
Eqs. (\ref{A1f}) and (\ref{A1fo}) can be used to determine the optimal
orientation of magnetic field for the observation of the beat frequency. For
straight interlayer electron coupling as in Eq. (\ref{kphim}), the beat
frequency has maximum value when magnetic field is perpendicular to the
layers, i.e. at polar angle $\theta =0$ . However, for the $\phi $-dependent
interlayer coupling at $\theta =0$ the beat frequency is zero, as follows
from Eq. (\ref{A1fo}). Hence, in this case to observe the beat frequency one
needs to incline the magnetic field. The angular dependence of the beat
frequency is given by the function in the curly brackets in Eq. (\ref{A1fo}%
). The first term $J_{m}\left( \kappa \right) \sin \left( m\varphi \right) $
in the curly brackets is much larger than the second. Its maximum gives the
optimal orientation ($\theta _{opt},\varphi _{opt}$) of magnetic field for
the observation of MQO beat frequency. For $m=2$, as in Sr$_{2}$RuO$_{4}$, Tl%
$_{2}$Ba$_{2}$CuO$_{6+\delta }$ and some other high-Tc compounds, the factor
$J_{m}\left( \kappa \right) \sin \left( m\varphi \right) $ has maximum at $%
\varphi _{opt}=(2n+1)\pi /4$ and $\theta _{opt}\approx \arctan \left(
3.0/k_{F}c^{\ast }\right) $, where $c^{\ast }$ is the interlayer lattice
constant. Note, that the spin factor of MQO also depends on the angle $%
\theta $.\cite{Shoenberg}
\begin{figure}[hht]
\includegraphics[width=0.49\textwidth]{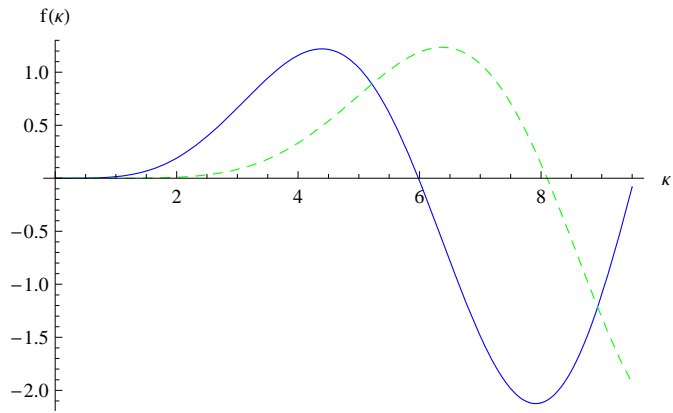}
\caption{{}(Color online) The amplitude of the $\varphi
$-dependence of the extremal cross-section area as function of the
polar angle $\theta $ for tetragonal (blue solid line) and
hexagonal (green dashed line) symmetries in the case of straight
interlayer electron hopping.} \label{FigPhiMax}
\end{figure}

If the beat frequency of MQO cannot be resolved (in dirty materials or at
high temperature), the minima of the beat frequency, i.e. the Yamaji angles,
can be detected from the increase of the amplitude of MQO. This increase of
MQO amplitude happens because at the Yamaji angles the MQO from both extremal
electron orbits have the same phase.\cite{Shoenberg,MarkReview} The Yamaji
angles can be much easier distinguished from the angular dependence of
background magnetoresistance (AMRO) (see Sec. IV). To determine the $\varphi $%
-dependence of the Yamaji angles one can again use Eqs. (\ref{A1f}) and (\ref%
{A1fo}).

If the $\varphi $-dependence of MQO beat frequency is clearly resolved, one
can obtain the information about the in-plane FS. Eqs. (\ref{A1f}) and (\ref%
{A1fo}) again can be used to determine the optimal magnetic field
orientation for the observation of this $\varphi $-dependence. In the case
of straight interlayer electron hopping, this $\varphi $-dependence $\propto
\cos m\varphi $ has the maximum amplitude when the factor $f\left( \kappa
\right) =\left( 1+\beta _{1}/\beta +m\right) J_{m}\left( \kappa \right)
-\kappa J_{m+1}\left( \kappa \right) $ in Eq. (\ref{A1f}) has maximum. For
typical value $\beta _{1}/\beta =1$ this factor as function of $\kappa
\equiv k_{F}c^{\ast }\tan \theta $ for $m=4$ and $m=6$ is plotted in Fig. %
\ref{FigPhiMax}. The function $f\left( \kappa \right) $ has first maximum at
$\kappa \approx m$ for $m=4$ and $m=6$ (see Fig. \ref{FigPhiMax}). It is
reasonable to use only the first maximum, because at high tilt angle of
magnetic field the cyclotron mass is large and the amplitude of MQO is too
small.

In the case of straight hopping, already the lowest-order harmonic in the $%
\varphi $-dependence of the MQO frequency gives the relative amplitude $%
\beta $\ of the same harmonic in the $\phi $-dependence of the in-plane
Fermi momentum [see Eq. (\ref{A1f})]. In the case of $\varphi $-dependent
interlayer electron hopping, given by Eq. (\ref{ko0}), in the main order,
the $\varphi $-dependence of MQO frequency comes from the $\phi $-dependence
of the interlayer transfer integral and does not give information about the
in-plane FS. To determine the shape of the in-plane FS, one needs to study
higher harmonics $\propto \cos \left( 3m\varphi \right) $ in the MQO
frequency. The amplitude of the $\cos \left( 3m\varphi \right) $ term in MQO
frequency is given by the function $f_{1}\left( \kappa \right) =\left(
1+\beta _{1}/\beta +3m\right) J_{3m}\left( \kappa \right) -\kappa
J_{3m+1}\left( \kappa \right) $ in Eq. (\ref{A1fo}) and has first maximum at
$\kappa \approx 3m$. This determines the optimal polar angle $\theta _{opt}$%
, at which this $\varphi $-dependence is most easily observed. According to
Eq. (\ref{A1fo}), this dependence gives the amplitude of the $2m$ harmonic
modulation of the in-plane FS.

\section{Elliptic Fermi surface}

Now we derive the analytical formula for the Yamaji angles for the elliptic
in-plane dispersion
\begin{equation}
\varepsilon \left( k_{x},k_{y}\right) \equiv
k_{x}^{2}/2m_{x}+k_{y}^{2}/2m_{y}=\varepsilon \left( k\right) \left[ 1+\beta
\cos 2\phi \right] ,  \label{Eelliptic}
\end{equation}%
where $\varepsilon \left( k\right) =k^{2}/2m,$ $m\equiv 2m_{x}m_{y}/\left(
m_{x}+m_{y}\right) $ and $\beta \equiv \left( m_{x}-m_{y}\right) /\left(
m_{x}+m_{y}\right) $. The shape of the FS for this dispersion is, of cause,
also elliptical. The ellipse can be obtained from the circle by the dilation
$\Lambda _{x}$ along one in-plane direction (along the $x$-axis): $%
x\rightarrow \lambda x$. Consider the cross-section area of the FS by the
plane, cutting the $k_{z}$-axis at the point $k_{z0}$, and perpendicular to
the magnetic field direction $\mathbf{B}=B\mathbf{n}$, where the unit vector
\begin{equation}
\mathbf{n}=\left( n_{x},n_{y},n_{z}\right) =\left( \sin \theta \cos \varphi
,\sin \theta \sin \varphi ,\cos \theta \right) .
\end{equation}%
For the circular in-plane FS this cross-section area is independent of the
angle $\varphi $. In the first order in $t_{c}/E_{F}$, it is also
independent of $k_{z0}$ at special directions $\mathbf{n}_{Yam},$
corresponding to the Yamaji angles $\theta =\theta _{Yam}$, given by Eq. (%
\ref{YamZero}). After the dilation $\Lambda _{x}$, the direction of magnetic
field, which is perpendicular to the cross-section plane, also changes:
\begin{equation}
\mathbf{n}\,\rightarrow \Lambda _{x}\left( \mathbf{n}\right) =\frac{\left(
n_{x}/\lambda ,n_{y},n_{z}\right) }{\sqrt{\left( n_{x}/\lambda \right)
^{2}+n_{y}^{2}+n_{z}^{2}}}.
\end{equation}%
However, the cross-section area perpendicular to $\,\mathbf{n}_{1}=\Lambda
_{x}\left( \mathbf{n}_{Yam}\right) $ remains independent of $k_{z0}$, if it
was independent before the dilation. Hence, the direction $\,\mathbf{n}%
_{1}=\Lambda _{x}\left( \mathbf{n}_{Yam}\right) $ corresponds to the new
Yamaji angle $\theta _{Yam}\left( \varphi \right) $. The polar and azimuthal
angles are related to the components of the vector $\mathbf{n}_{1}=\left(
n_{1x},n_{1y},n_{1z}\right) $ as%
\begin{equation}
\tan \theta _{1}=\frac{\sqrt{n_{1x}^{2}+n_{1y}^{2}}}{n_{1z}},~\tan \varphi
_{1}=\frac{n_{1y}}{n_{1x}}.
\end{equation}%
Combining above equations we obtain the relation between the old and new
Yamaji angle $\theta _{Yam}^{\ast }=\Lambda _{x}\left( \theta _{Yam}\right) $%
\begin{equation*}
\frac{\tan \theta _{Yam}^{\ast }}{\tan \theta _{Yam}}=\frac{\sqrt{%
n_{x}^{2}/\lambda ^{2}+n_{y}^{2}}}{n_{z}\tan \theta _{Yam}}=\sqrt{\frac{\cos
^{2}\varphi }{\lambda ^{2}}+\sin ^{2}\varphi }.
\end{equation*}%
The angle $\varphi $ here is the angle before the dilation $\Lambda _{x}$.
It is related to the angle $\varphi _{1}$ after the dilation as
\begin{equation*}
\tan \varphi _{1}=\lambda \tan \varphi .
\end{equation*}%
Then, after simple trigonometric algebra, we obtain
\begin{equation}
\frac{\tan \theta _{Yam}^{\ast }}{\tan \theta _{Yam}}=\frac{\cos \varphi }{%
\lambda }\sqrt{1+\tan ^{2}\varphi _{1}}=\frac{\sqrt{1+\tan ^{2}\varphi _{1}}%
}{\sqrt{\lambda ^{2}+\tan ^{2}\varphi _{1}}}=\frac{1}{\sqrt{\lambda ^{2}\cos
^{2}\varphi _{1}+\sin ^{2}\varphi _{1}}}.  \label{TYam}
\end{equation}%
For the elliptic dispersion (\ref{Eelliptic}) the maximum value of the Fermi
momentum projection on the in-plane magnetic field direction is given by
\begin{equation}
p_{B}^{\max }=\sqrt{\left( p_{1}\cos \varphi \right) ^{2}+\left( p_{2}\sin
\varphi \right) ^{2}},  \label{pbell}
\end{equation}%
where $p_{1}^{2}=2m_{x}\varepsilon _{F}$ and $p_{2}^{2}=2m_{y}\varepsilon
_{F}$. The r.h.s. of Eq. (\ref{TYam}) coincides with $p_{2}/p_{B}^{\max }$.
Hence, the generalization of the Yamaji zeros to the elliptic dispersion (%
\ref{Eelliptic}) writes down as%
\begin{equation}
J_{0}\left[ c^{\ast }p_{B}^{\max }\left( \varphi \right) \tan \theta _{n}%
\right] =0.  \label{YamZerosPhi}
\end{equation}

Approximately, Eq. (\ref{YamZerosPhi}) coincides with Eq. (\ref{DThetaS}),
derived for the interlayer conductivity\cite{Mark92,MarkReview} from the
Shockley tube integral\cite{Ziman}. The saddle point approximation, used in
Ref. \cite{Mark92} to derive Eq. (\ref{DThetaS}), assumes that the $z$%
-component of electron velocity oscillates rapidly when the
electron moves along its closed orbit in the momentum space. This
is valid only at high tilt angles $\theta $ of magnetic field,
when $\tan \theta \gg 1/c^{\ast }p_{B}^{\max }\left( \varphi
\right) $, and only in very clean samples with $ \omega _{c}\tau
\cos \theta \gg 1$. The reason, why Eq. (\ref{DThetaS}) describes
well some experimental data,\cite{MarkReview} comes from its
coincidence with the exact geometrical expression
(\ref{YamZerosPhi}) for the Yamaji angles for the elliptic Fermi
surface, which according to Eq. (\ref{SYamaji}) gives the maxima
of magnetoresistance.

\section{Magnetoresistance}

To calculate magnetoresistance as function of the direction $\left( \theta
,\varphi \right) $ of magnetic field one can use the quasi-classical
Boltzmann transport equation for electrons moving along the closed orbits in
magnetic field. This approach gives the Shockley-Chambers formula\cite{Ziman}%
, which at zero temperature expresses conductivity tensor $\sigma _{\alpha
\beta }$ via the integral over the Fermi surface:
\begin{eqnarray}
\sigma _{\alpha \beta }\left( \theta ,\varphi \right) &=&\frac{e^{2}}{4\pi
^{3}\hbar ^{2}}\int dk_{z0}\frac{m_{H}^{\ast }\cos \theta \,/\omega _{H}}{%
1-\exp \left( -2\pi /\omega _{H}\tau \right) }  \label{ST} \\
&&\times \int_{0}^{2\pi }\int_{0}^{2\pi }v_{\alpha }\left( \psi
,k_{z0}\right) v_{\beta }\left( \psi -\psi ^{\prime },k_{z0}\right) e^{-\psi
^{\prime }/\omega _{H}\tau }d\psi ^{\prime }d\psi .  \notag
\end{eqnarray}%
Here the momentum space is parametrized by the momentum component $%
k_{H}=k_{z0}\cos \theta $ along the magnetic field, by energy $E$ and by the
angle $0<\psi <2\pi $ of the rotation in the cross-section plane. The
effective cyclotron mass of the orbit is given by%
\begin{equation}
m_{H}^{\ast }\equiv \frac{1}{2\pi }\frac{\partial A}{\partial E}=\frac{1}{%
2\pi \varepsilon ^{\prime }\left( k_{F}\right) }\frac{\partial A}{\partial
k_{F}},  \label{mH1}
\end{equation}%
where $A=A\left( k_{H},E_{F}\right) $ is the area of the FS cross-section
perpendicular to the magnetic field $\mathbf{B}$ at the momentum $%
k_{H}\parallel \mathbf{B}$. The cyclotron frequency of the orbit $\omega
_{H}\equiv eB/m_{H}^{\ast }c$, and $v_{\alpha }\left( \psi ,k_{H}\right) $
is the component of the electron velocity on the FS. Generally, the mean
scattering time $\tau $ in the integrand (\ref{ST}) may also depend on the
position on the Fermi surface. However, we neglect this dependence because
in the simplest theory of spin-independent short-range impurity scattering $%
\tau $ depends only on the density of states at the Fermi level.

For dispersion (\ref{1}) the electron velocity component along the $z$-axis
is a function of $k_{z}$ only:
\begin{equation}
v_{z}\left( k_{z}\right) =\left( 2c^{\ast }t_{z}/\hbar \right) \sin \left(
c^{\ast }k_{z}\right) .  \label{vz}
\end{equation}%
The $k_{z}$ coordinate of the FS point $K$ satisfies the equation (\ref%
{kappa}), where $\varphi +\phi ^{\prime }$ is the azimuthal angle of the
projection of the FS point $K$ on the $x$-$y$ plane. Approximately, this
equation can be solved by the iteration procedure. In the zeroth order
\begin{equation}
k_{z}^{(0)}=k_{z0}-k_{F}\cos \left( \phi ^{\prime }\right) \tan \theta ,
\label{kz0}
\end{equation}%
and in the next orders%
\begin{equation}
k_{z}^{(i+1)}=k_{z0}-k_{F}\left( \phi ^{\prime }+\varphi ,k_{z}^{(i)}\right)
\cos \left( \phi ^{\prime }\right) \tan \theta .  \label{kz1}
\end{equation}

The angle $\psi $ entering the Shockley-Chambers formula (\ref{ST})
corresponds to the increment of the cross-section area at a given increment
of energy:
\begin{equation*}
d\psi =\frac{1}{m_{H}^{\ast }}\frac{dk}{v_{\perp }}=\frac{dk\,\partial
k_{\perp }}{m_{H}^{\ast }\partial E}.
\end{equation*}%
Generally, $\psi $ is different from the angles $\phi ^{\prime }$ and $\phi $
of the rotation in the cross-section and in the $x$-$y$ planes. The
cross-section area multiplied by $\cos \theta $ is equal to the area of the
projection in the $x$-$y$ plane, and $\psi $ is related to the angle $\phi $
of the rotation in the $x$-$y$ plane as%
\begin{equation}
\frac{d\psi }{d\phi }=\frac{k_{F}\left( \phi ,k_{z}\right) }{m_{H}^{\ast
}\cos \theta }\frac{\partial k_{F}\left( \phi ,E\right) }{\partial E}.
\label{psi}
\end{equation}%
For cylindrical FS one has $k_{F}\left( \phi \right) =k_{F}$, and $\psi $
coincides with the angle $\phi $.

Now we show that for $\omega _{H}\tau \gg 1$ the minima of $\sigma
_{zz}\left( \theta ,\varphi \right) $, given by Eq. (\ref{ST}), coincide
with the minima of the mean-square value of the derivative $\left( \partial
A/\partial k_{z0}\right) $, i.e. with the geometrical Yamaji angles. At $%
\omega _{H}\tau \gg 1$ the exponent $e^{-\psi ^{\prime }/\omega _{H}\tau
}\approx 1$, and Eq. (\ref{ST}) gives%
\begin{equation}
\sigma _{\alpha \alpha }\left( \theta ,\varphi \right) =\frac{e^{2}}{4\pi
^{3}\hbar ^{2}}\int dk_{z0}\frac{m_{H}^{\ast }\cos \theta \,/\omega _{H}}{%
1-\exp \left( -2\pi /\omega _{H}\tau \right) }\left( \int_{0}^{2\pi
}v_{\alpha }\left( \psi ,k_{z0}\right) d\psi \right) ^{2}.  \label{ssimple}
\end{equation}%
Using Eq. (\ref{psi}) we transform the integral
\begin{eqnarray}
I &\equiv &\int_{0}^{2\pi }d\psi v_{z}\left( \psi ,k_{z0}\right)  \notag \\
&=&\int_{0}^{2\pi }d\phi \frac{k_{F}\left( \phi ,k_{z}\right) }{m_{H}^{\ast
}\cos \theta }\frac{\partial k_{F}\left( \phi ,E\right) }{\partial E}\frac{%
\partial E}{\partial k_{z}}  \notag \\
&=&\int_{0}^{2\pi }d\phi \frac{k_{F}\left( \phi ,k_{z}\right) }{m_{H}^{\ast
}\cos \theta }\frac{\partial k_{F}\left( \phi ,k_{z}\right) }{\partial k_{z}}%
.  \label{I}
\end{eqnarray}%
The derivative%
\begin{equation}
\frac{\partial k_{F}\left( \phi ,k_{z}\right) }{\partial k_{z}}=\frac{%
\partial k_{F}\left[ \phi ,k_{z}\left( k_{z0},\phi \right) \right] }{%
\partial k_{z0}\cdot \left( \partial k_{z}/\partial k_{z0}\right) }.
\end{equation}%
From (\ref{kappa}) in the first order in $t_{z}$ we obtain $\partial
k_{z}/\partial k_{z0}=1$. Hence, from (\ref{I}) we get%
\begin{equation*}
I=\int_{0}^{2\pi }\frac{d\phi }{m_{H}^{\ast }\cos \theta }\frac{\partial
k_{F}^{2}\left( \phi ,k_{z}\right) }{2\partial k_{z0}}=\frac{\partial
A\left( k_{z0},\theta ,\varphi _{0}\right) }{\partial k_{z0}~m_{H}^{\ast }},
\end{equation*}%
where the cross-section area $A$ is given by Eq. (\ref{A}).

Now from (\ref{ssimple}) we get
\begin{equation}
\sigma _{zz}\left( \theta ,\varphi \right) =\frac{e^{2}\tau \cos \theta }{%
8\pi ^{4}\hbar ^{2}}\int \frac{dk_{z0}}{m_{H}^{\ast }}\left( \frac{\partial
A\left( k_{z0},\theta ,\varphi \right) }{\partial k_{z0}~}\right) ^{2}.
\label{SYamaji}
\end{equation}%
Similar relation without rigorous proof was also proposed in Ref. \cite%
{Mark92}. Eq. (\ref{SYamaji}) means, that the angular dependence of the
interlayer conductivity $\sigma _{zz}$ and of the mean-squared derivative of
the FS cross-section area $\partial A/\partial k_{z0}$ coincide in the limit
$\omega _{H}\tau \gg 1$. In particular, the geometrical Yamaji angles
coincide with the minima of interlayer conductivity at $\omega _{H}\tau \gg
1 $.

\section{Summary and discussion}

Above we have obtained the following main results: (i) the analytical
formulas (\ref{A1f}),(\ref{A1fo}) for the main $\varphi $-dependent term of
the cross-section area $A\left( k_{z0},\theta ,\varphi \right) $, when the
FS corrugation is weak (Sec. III); (ii) the exact analytical formula (\ref{DThetaS}) for the Yamaji angles in the case of elliptic in-plane Fermi surface (Sec.
IV); (iii) the derivation of Eq. (\ref{SYamaji}), which states that in the
limit $\omega _{H}\tau \gg 1$ the angular oscillations of magnetoresistance
coincide with the angular oscillations of the $k_{z}$-dependent term in the
cross-section area. Eq. (\ref{SYamaji}) brings additional importance to the
results in Secs. III and IV for the FS cross-section area. In particular,
Eq. (\ref{SYamaji}) means, that the geometrical Yamaji angles $\theta
_{Yam}\left( \varphi \right) $ coincide with the maxima of magnetoresistance
at $\omega _{H}\tau \gg 1$.

If the interlayer electron hopping is $\phi $-dependent, as in Eq. (\ref{ko0}%
), Eq. (\ref{SYamaji}) also suggests the very strong $\varphi $-dependence
of the interlayer magnetoresistance, given by Eq. (\ref{A1fo}). Note, that
this $\varphi $-dependence of magnetoresistance for Sr$_{2}$RuO$_{4}$ and Tl$%
_{2}$Ba$_{2}$CuO$_{6+\delta }$ is in contrast to the so-called "third
angular effect", developed in Ref. \cite{LebedBagmet} for the in-plane
magnetic field direction. For the Fermi surface in Fig. \ref{FigTetr}a, the
third angular effect\cite{LebedBagmet} predicts maxima of conductivity at $%
\varphi =0$, coinciding with the positions of the FS inflection points,
while Eq. (\ref{A1fo}) predicts these maxima at $\varphi =\pi /4$.

The $\theta -\varphi $ dependence of the cross-section area is
also an important result, because the magnetic quantum
oscillations give the extremal cross-section area as function of
magnetic field orientation. In the end of Sec III we summarized,
how the formulas (\ref{A1f}) and (\ref{A1fo}) for the
cross-section area can be applied to analyze MQO. In particular,
these formulas give (i) the optimal direction of magnetic field
for the observation of the beats of MQO and of the $\varphi
$-dependence of the beat frequency; (ii) the Yamaji angles, where
the magnetoresistance and the amplitude of MQO have maxima; (iii)
the relation between the FS shape and the $\varphi $-dependence of
the beat frequency.

Now we discuss in more details the applicability region of all
above results and compare them with the previous theoretical
results. Let us first study the analytical formula (\ref{A1f}) for
the cross-section area, obtained from the Fourier expansion in
the limit of weak FS warping. This formula strongly differs from
the previous result\cite{Bergemann} [see Eq. (\ref{A1})]. The
derivation of the result of Bergemann et al.\cite{Bergemann} does
not include the second-order terms in the FS corrugation $k_{\mu
\nu }/k_{00} $, which is necessary, because for the straight
interlayer electron hopping and in the first order in the FS
corrugation, the $\varphi $-dependence of the FS cross section
disappears. The differences between the result of Ref.
\cite{Bergemann} and the new formula (\ref{A1f}) are illustrated
in Figs. \ref{FigMon},\ref{FigTetr} Roughly, the formula of
Bergemann et al.\cite{Bergemann} gives too small $\varphi
$-dependence of the cross-section area, which is smaller than the
correct result, approximately, by a factor of $2+m$, where $m$ is the harmonic index of this $\varphi $-dependence.

Eqs. (\ref{A1f}) and (\ref{A1fo}), derived using the
small harmonic expansion, turns out to be a very good
approximation for any typical FS with the tetragonal or hexagonal
symmetry (see Fig. \ref{FigTetr}). This is because even a small
value of $\beta $ leads to the strong change of the in-plane FS
shape: for the almost quadratic in-plane FS as in Fig.
\ref{FigTetr}a, $\left\vert \beta \right\vert =0.06\ll 1$. Eq.
(\ref{A1f}) may considerably violate only in the case of
monoclinic or triclinic symmetry, when the in-plane FS is strongly
elongated (see Fig. \ref{FigMon}a). Fortunately, just for this
case Eq. (\ref{YamZerosPhi}) gives the reliable result (see Fig.
\ref{FigMon}b).

The $\theta $-$\varphi $ dependence of the cross-section area,
given by Eqs. (\ref{A1f}) and (\ref{A1fo}), allows to extract not
only the leading in-plane $\varphi $-dependence of the FS, but
also to get some information about the electron Fermi velocity 
from the coefficient $\beta _{1}$. However, the extraction of $\beta _{1}$ is much more difficult than the extraction of $\beta $, because,
according to Eqs. (\ref{A1f}) and (\ref{A1fo}), the dependence of
the cross-section area $A_{1}\left( \theta ,\varphi \right) $ on
$\beta _{1}$ is much weaker than on $\beta $. In Figs.
\ref{FigMon}b,\ref{FigTetr}b, for definiteness, we take the
dispersion of the form (\ref{e1}) and $t_{c}\left( \phi \right) $
to be independent on $\phi $, which gives Eq. (\ref{kF1s}) and
$\beta _{1}/\beta =1$, $C_{1}=1$ in Eqs. (\ref{A1f}) and (\ref{A1fo}).

The $\varphi $-dependence of the Yamaji angles turns out to be weak in the
cases of tetragonal or hexagonal symmetry. This dependence is much weaker
than the prediction of Eq. (\ref{YamZerosPhi}), used in Refs. \cite{Mark92,Nam1,HousePRB1996,BanguraPRB2007} for the elongated FS, and is much stronger than the prediction of Eq. (\ref{A1}), used in Refs. \cite{Bergemann}. For the first Yamaji angle in the case of tetragonal symmetry,
Eq. (\ref{YamZerosPhi}) even gives the opposite sign of the $\varphi $%
-dependence of the first Yamaji angle, as can be seeing from Fig. \ref%
{FigTetr}b. For the superelliptic FS, given by Eq. (\ref{SupElDisp}) and
discussed in Ref. \cite{Nam1}, there is a strong tetragonal modulation of
the elliptic FS, and Eq. (\ref{YamZerosPhi}) also fails to give a reliable
result.

Eqs. (\ref{A1f}),(\ref{A1fo}),(\ref{YamZerosPhi}),(\ref{SYamaji})
are valid only in the first order in the small parameter $t_{\perp
}/\varepsilon _{F}$. This is, usually, a good approximation for
the layered high-Tc superconductors, organic metals, and many
other compounds. However, some fine details of the angular
dependence of magnetoresistance may be sensitive to the next
interlayer hopping term, especially in the case, when the main
interlayer hopping is strongly $\phi $-dependent.

\begin{figure}[tbh]
\includegraphics[width=0.49\textwidth]{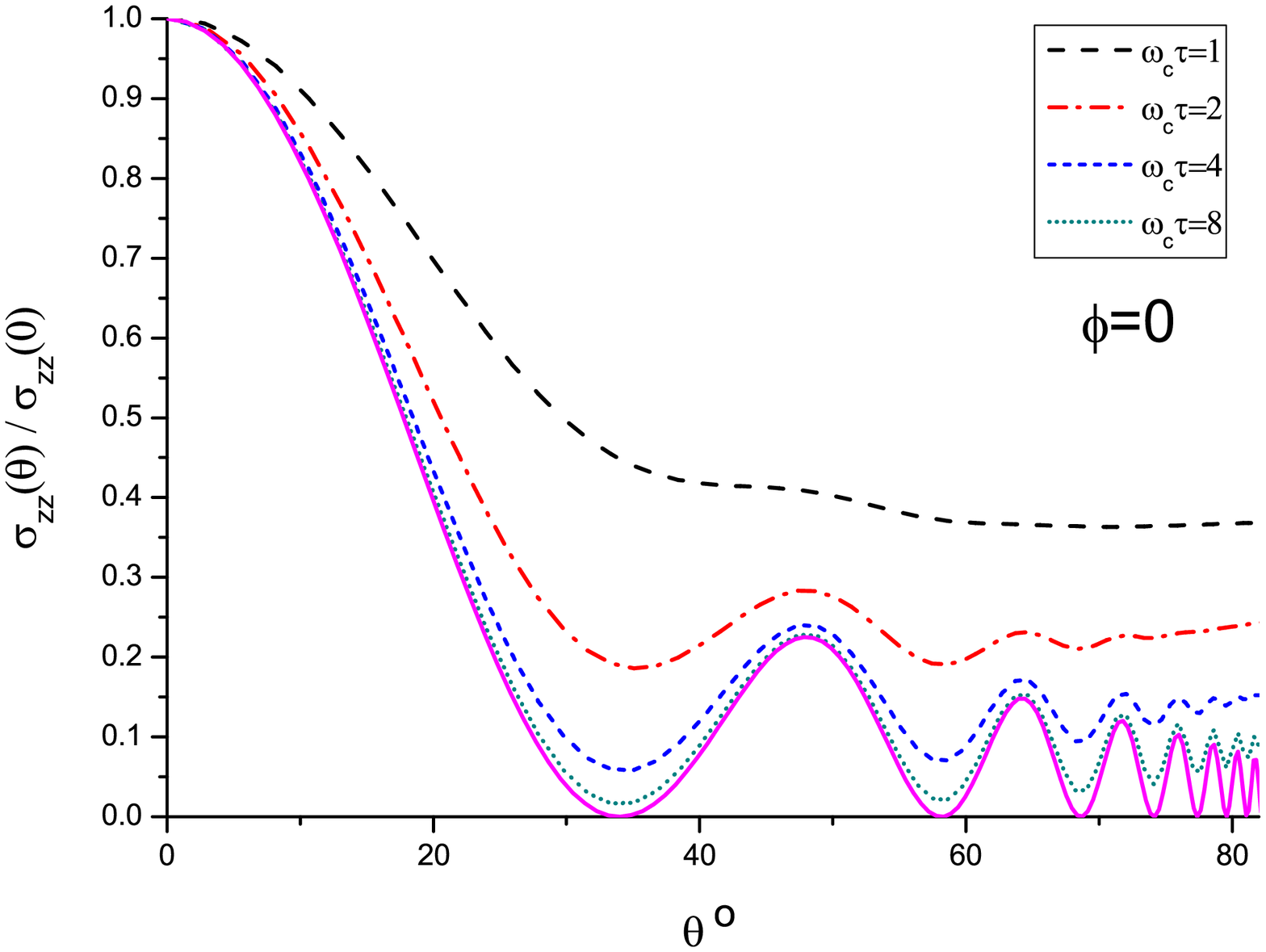}a\newline
\includegraphics[width=0.49\textwidth]{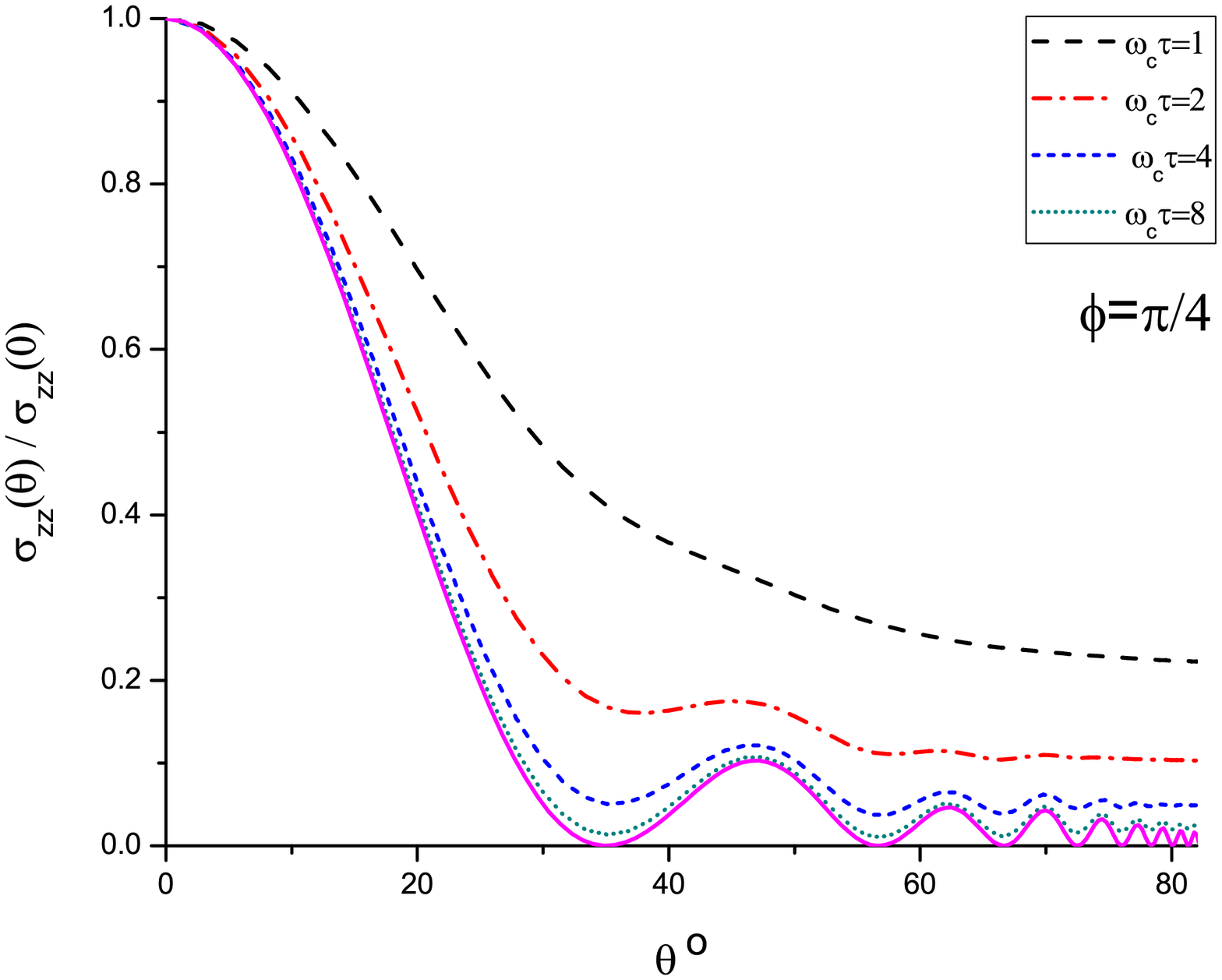}b
\caption{{}The normalized conductivity $\protect\sigma _{zz}$, calculated
from Eq. (\protect\ref{ST}) for the dispersion (\protect\ref{DispHTc}) as
function of the polar angle $\protect\theta $ for two different azimuthal
angles $\protect\phi $ at several values of $\protect\omega _{c}\protect\tau %
=1$ (black dashed line), $\protect\omega _{c}\protect\tau =2$ (red
dash-dotted line), $\protect\omega _{c}\protect\tau =4$ (blue short-dashed
line) and $\protect\omega _{c}\protect\tau =8$ (green dotted line). The
solid magenta line gives the result of Eq. (\protect\ref{SYamaji}).}
\label{FigST}
\end{figure}

Eqs. (\ref{A1f}),(\ref{A1fo}),(\ref{YamZerosPhi}) determine the
geometrical conditions for the FS cross section to be almost
independent on the interplane momentum $k_{z}$, which, according
to Eq. (\ref{SYamaji}), gives the minima of conductivity at
$\omega _{c}\tau \rightarrow \infty $. To check how strong the
conductivity, given by the Shockley-Chambers formula (\ref{ST}),
differs from the geometrical formula (\ref{SYamaji}) at finite
$\omega _{c}\tau $, we perform the numerical calculation of the
$\theta $-dependence of conductivity, $\sigma _{zz}\left( \theta
,\varphi ,\omega _{c}\tau \right) $, given by Eq. (\ref{ST}), at
four different values of $\omega _{c}\tau =1,2,4,8$ and for two
values of the azimuthal angle, $ \varphi =0$ and $\varphi =\pi
/4$. Then, we compare it with the dependence $ \sigma _{zz}\left(
\theta ,\varphi \right) $ given by Eq. (\ref{SYamaji}). The
results for the polar-angle dependence of the normalized
interlayer conductivity are given in Figs. \ref{FigST}. In this
calculation we take the in-plane dispersion, proposed for the
layered cuprate high-Tc superconductors and given by\cite{HusseyNature2003,AbdelPRL2007AMRO,McKenzie2007}
\begin{equation}
\varepsilon \left( k_{x},k_{y}\right) =2t_{1}\left[ \cos \left(
k_{x}a\right) +\cos \left( k_{y}a\right) \right] +4t_{2}\cos \left(
k_{x}a\right) \cos \left( k_{y}a\right) -2t_{3}\left[ \cos \left(
2k_{x}a\right) +\cos \left( 2k_{y}a\right) \right] -E_{F},  \label{DispHTc}
\end{equation}%
where $a=3.95\mathring{A}$ is the lattice constant, $%
t_{1}=0.38eV,~t_{2}=0.32t_{1},~t_{3}=0.5t_{2}$. We take the
doping-dependent Fermi energy $E_{F}=0.02123eV$. The FS for this
dispersion (\ref{DispHTc}) has tetragonal symmetry and is very
similar to that in Fig. \ref{FigTetr}a
with slightly different values $k_{F}a\approx 2.14$ at $\varphi =0$ and $%
k_{F}a\approx 2.43$ at $\varphi =\pi /4$. Therefore, we don't plot it again.
The interlayer hopping term of electron dispersion is taken to be straight ($%
\phi $-independent) and given by Eq. (\ref{kphim}). From Fig. \ref{FigST} we
see, that the geometrical formula (\ref{SYamaji}) gives rather accurate
results for the Yamaji angles of magnetoresistance at $\omega _{c}\tau
\gtrsim 2$. For large values of $\omega _{c}\tau $, the results of Eqs. (\ref%
{SYamaji}) and (\ref{ST}) coincides. At $\omega _{c}\tau \lesssim 1$\ the
difference between Eqs. (\ref{SYamaji}) and (\ref{ST}) for the first Yamaji
angle reaches 5-10\%. The $\varphi $-dependence of the Yamaji angles given
by Eqs. (\ref{SYamaji}) and (\ref{ST}) agrees very well; only the amplitude
of AMRO reduces with decreasing $\omega _{c}\tau $.

The background magnetoresistance and, in particular, the
saturation values of $\sigma _{zz}$ at $\theta \rightarrow
90^{\circ }$, depend strongly on the azimuthal angle $\varphi $
(see Fig. \ref{FigST}). Therefore, it is reasonable to use this
$\varphi $-dependence of the conductivity saturation value at
$\theta =\pi /2$ to determine the in-plane Fermi surface from the
experimental data on the angular dependence of magnetoresistance.
The theoretical prediction for this dependence can be obtained
from the numerical calculation using the Shockley-Chambers formula
(\ref{ST}), as is done in Fig. \ref{FigST}. The origin of this
$\varphi $-dependence is qualitatively explained in Ref.
\cite{LebedBagmet}.

\section{Acknowledgment}

The work was supported by MK-2320.2009.2, by RFBR 09-02-12206-OFI-M and by 
the Foundation "Dynasty".

\end{document}